\documentclass{article}
\usepackage{spconf,amsmath,graphicx}
\usepackage{array}
\usepackage{algorithm}
\usepackage{algorithmic}
\usepackage{rotating}
\usepackage{hyperref}
\usepackage{multirow}
\usepackage{booktabs}
\usepackage{longtable}

\title{Multi-branch learning for weakly-labeled Sound event detection}
\name{Yuxin Huang$^{1,2}$, Xiangdong Wang$^{1,\dagger}$, Liwei Lin$^{1,2}$, Hong Liu$^1$, Yueliang Qian$^1$}
\address{
  $^1$Beijing Key Laboratory of Mobile Computing and Pervasive Device,\\
  Institute of Computing Technology,
  Chinese Academy of Sciences, Beijing, China\\
  $^2$University of Chinese Academy of Sciences, Beijing, China\\
  \{huangyuxin18g, xdwang, linliwei17g, hliu, ylqian\}@ict.ac.cn}
\begin{document}
%
\maketitle
\begin{abstract}
There are two sub-tasks implied in the weakly-supervised SED: audio tagging and event boundary detection. Current methods which combine multi-task learning with SED requires annotations both for these two sub-tasks. Since there are only annotations for audio tagging available in weakly-supervised SED, we design multiple branches with different learning purposes instead of pursuing multiple tasks. Similar to multiple tasks, multiple different learning purposes can also prevent the common feature which the multiple branches share from overfitting to any one of the learning purposes. We design these multiple different learning purposes based on combinations of different MIL strategies and different pooling methods. Experiments on the DCASE 2018 Task 4 dataset and the URBAN-SED dataset both show that our method achieves competitive performance.

\end{abstract}
\begin{keywords}
Sound event detection, weakly-labeled, multi-branch learning, multi-task learning
\end{keywords}
\section{Introduction}
\label{sec:intro}

Sound event detection (SED) is a task that aims to detect acoustic events in an audio recording. It not only judges whether a specific sound event exists in the audio but also estimates the onset and offset of the existing sound event.

Currently, due to the excellent performance of neural networks, most SED systems are based on neural networks \cite{cakir2017convolutional,adavanne2017sound}.
Especially, Phan et al. \cite{phan2019unifying} and Xia et al. \cite{xia2019multi} propose methods for SED combining multi-task learning and neural networks, in which audio tagging and event boundary detection are identified as two separate tasks.
There are two kinds of multi-task learning methods: hard parameter sharing and soft parameter sharing multi-task learning.
For the hard parameter sharing methods, hidden layers are shared among multi-tasks and specific output layers are kept for each task \cite{caruana1997multitask, ruder2017overview}. As a result, the feature representations generated by the shared hidden layers can be applied to the multiple tasks, which reduces the risk of overfitting \cite{ruder2017overview, baxter1997bayesian}.
Soft parameter sharing multi-task learning regularizes the distance between the parameters of each task, which makes the parameters of each task similar \cite{ruder2017overview}.
The multi-task learning methods require labels for each task to learn. Since most multi-task methods for SED regard audio tagging and event boundary detection as two separate tasks, strong labels containing the two following annotations are required: occurrences of event categories in an audio clip and the onset and offset of each event category. However, due to the high cost of manual annotations \cite{mesaros2016tut},
the strong labels are difficult to obtain. As a result, weakly-supervised SED, which utilizes weak annotations that only labeled with the occurrences of event categories, has become a new research focus.

Weakly-supervised SED is usually approached as a multiple instance learning (MIL) problem \cite{dietterich1997solving, pathak2014fully}.
According to MIL, a bag of instances has only the bag-level label. Then for weakly-supervised SED, an MIL strategy provides a way to decide the clip-level prediction of an audio clip depending on the frame-level information. The implementation of the combination of MIL and neural networks relies on a feature encoder and a pooling block. The feature encoder can be a neural network such as CNN, RNN and CRNN. The pooling block consists of a classifier and a pooling module which adopts pooling methods such as global max pooling (GMP) \cite{oquab2015object}, global average pooling (GAP) \cite{zhou2016learning}, global weighted rank pooling (GWRP) \cite{kolesnikov2016seed}, attention pooling (ATP) \cite{ilse2018attention,xu2018large} and so on. There are two types of MIL strategies \cite{ilse2018attention}: the instance-level approach and the embedding-level approach.
For the instance-level approach, the pooling module integrates the frame-level probabilities output by the classifier into a clip-level probability. For the embedding-level approach, the pooling module integrates the frame-level feature representations output by the feature encoder into a clip-level feature representation, which is passed through the classifier to get the clip-level probability.
As discussed in \cite{ilse2018attention}, the embedding-level approach tends to outperform the instance-level approach in terms of clip-level prediction. This is because the different learning purposes implied in the different MIL strategies lead to different characteristics that the feature serves. More specifically, the clip-level feature in the embedding-level approach enables the feature encoder to generate high-level feature representations with more clip-level characteristics. On the contrary, the usage of the frame-level feature in the instance-level approach leads to more frame-level characteristics.

In this paper, we propose a multi-branch learning (MBL) method which provides a way to combine multi-task learning with weakly-supervised SED.
The structure of MBL consists of a feature encoder and multiple branches. Different branches employ the same high-level feature generated by the feature encoder to predict the target.
Similar to multi-task learning, although all the branches of MBL predict the same eventual target, they focus on different characteristics of the common feature, which enables the common feature to be fit for various learning purposes and reduces the risk of overfitting single learning purpose. We design such two different types of branches inspired by the different learning purposes between the embedding-level approach and the instance-level approach.
Further more, by considering different pooling methods, one type of the branches is implemented as several auxiliary branches for training, while another type of the branches is implemented as a main branch both for training and detection. We regard the branch with the embedding-level approach as the main branch and those with the instance-level approach as auxiliary branches. Compared to multi-task learning for SED, our approach only requires weak labels.
Experiments on both the DCASE 2018 Task 4 dataset and the URBAN-SED dataset show that the proposed MBL achieves excellent performance.

\section{The MIL strategies for weakly-supervised SED}
\label{sec:releated_work}
In this section, we introduce how 3 typical pooling methods including global max pooling (GMP), global average pooling (GAP) and attention pooling (ATP) are implemented with the two MIL strategies (the embedding-level and instance-level approaches).

\begin{table}
\small
\caption{The overview of the three pooling methods }
\label{table1}
\begin{center}
\begin{tabular}{|c|l|l|}
\hline
Method & Instance-level & Embedding-level
\\ \hline
GMP & $\hat{\mathbf{P}}({y_c}|\mathbf{x}) = \max\limits_{t}{\mathbf{P}_{frame}({y_c}|{x_{t}})}$ &
$ h_c^e = \max\limits_{t}{x_t^e}$
\\ \hline
GAP & $\hat{\mathbf{P}}({y_c}|\mathbf{x}) =
 \frac{1}{T}\sum\limits_{t}{\mathbf{P}_{frame}({y_c}|{x_{t})}}$ &
 $h_c = \frac{1}{T} \sum\limits_{t}x_{t}$
\\  \hline
ATP & $\hat{\mathbf{P}}({y_c}|\mathbf{x}) = \sum\limits_{t}a_{ct} \mathbf{P}_{frame}(y_c|{x_t})$ &
$ {h_c} = \sum\limits_{t}{a_{ct} \cdot x_t}$
\\  \hline
\end{tabular}
\end{center}
\vskip -0.3in
\end{table}
\abovedisplayskip 0.07in
\belowdisplayskip 0.07in

Let $\mathbf{x} = \{x_1,...,x_T\}$ be the high-level feature representations of an audio clip generated by the feature encoder, and $y = \{y_1,...,y_C\} (y_c \in \{0,1\})$ be the ground truths, where $C$ is the number of event categories.

For the instance-level approach, the high-level feature representations are passed through the classifier to produce frame-level probabilities $\mathbf{P}_{frame}({y_c}|{x_{t}})$, then the pooling module integrates all the $\mathbf{P}_{frame}({y_c}|{x_{t}})$ into a clip-level probability $\hat{\mathbf{P}}({y_c}|\mathbf{x})$.
\begin{equation}
    \hat{\mathbf{P}}(y_c|\mathbf{x})=POOLING({\mathbf{P}_{frame}(y_c|x_1),...,\mathbf{P}_{frame}(y_c|x_T)})
\end{equation}

For the embedding-level approach, the pooling module transforms the high-level feature representations into a contextual representation $h_c$ at first:
\begin{equation}
h_c = POOLING(x_1, x_2,...,x_T)
\end{equation}
Then the contextual representation is passed through the classifier to produce the clip-level probability:
\begin{equation}
    \hat{\mathbf{P}}({y_c}|\mathbf{x}) = \mathbf{P}_{clip}({y_c}|h_c)
\end{equation}

Table \ref{table1} illustrates the details of the implementation of the 3 pooling methods with these 2 MIL strategies. Here, $h_c^e$ in the table represents the value of the $e^{th}$ dimension of $h_c$, and the weight $a_{ct}$ depends on trainable parameters $w_c$:
\begin{equation}
   a_{ct} = \frac{\exp((w_c^T x_t)/d)}
{\sum\limits_{k}{\exp((w_c^T x_k)/d)}}
\end{equation}
The hyper-parameter $d$ in ATP is used to scale the weight distribution appropriately.

The probability of the $t^{th}$-frame produced by the instance-level approach is:
\begin{equation}
   \hat{\mathbf{P}}(y_c|x_t) = \mathbf{P}_{frame}(y_c|x_t)
\end{equation}

However, there is no frame-level probability produced in the process of the embedding-level approach. To enable frame-level detection, we pass the frame-level high-level feature representations through the classifier to produce the frame-level probabilities:
\begin{equation}
 \hat{\mathbf{P}}(y_c|x_t) = \mathbf{P}_{clip}(y_c|x_t)
\end{equation}

Besides, for ATP, rather than using $\mathbf{P}_{clip}(y_c|x_t)$, we utilize the following frame-level probabilities:
\begin{equation}
    \hat{\mathbf{P}}(y_c|x_t) = Sigmoid{((w_c^T x_t)/d)}
\end{equation}

\section{Multi-branch learning}
\label{sec:model}

In this section, we describe in detail the proposed multi-branch learning (MBL) method for weakly-supervised SED.

The structure of MBL consists of a feature encoder and multiple branches.
Since we argue that similar to multiple tasks, multiple learning purposes can also prevent the feature encoder from overfitting, we design multiple branches with different learning purposes.

\begin{figure}[t]
\centering
\includegraphics[width=0.85\linewidth]{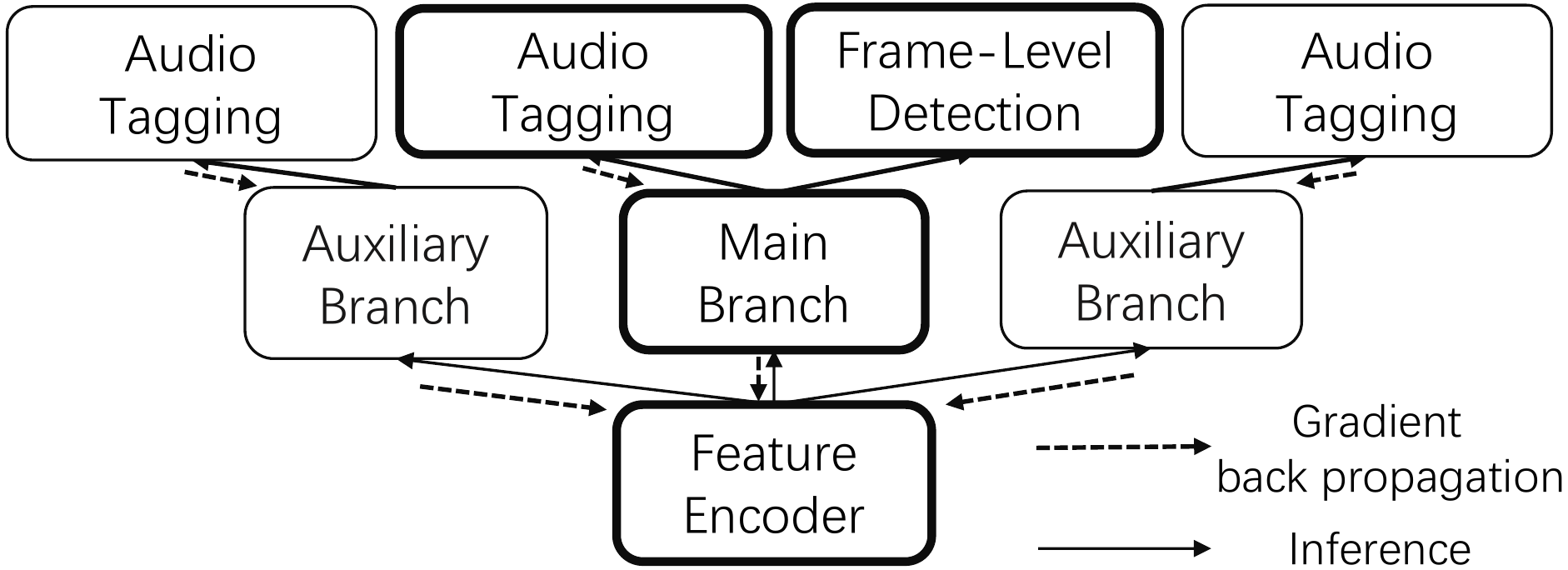}
\vskip -0.12in
\caption{The multi-branch learning method}
\label{fig2}
\vskip -0.21in
\end{figure}

For weakly-supervised SED, we follow the basic MIL framework mentioned in section \ref{sec:releated_work}. However, different from current methods, we employ multiple combinations of different MIL strategies and different pooling methods such as GAP and GMP to produce multiple branches with different learning purposes.
As shown in Figure~\ref{fig2}, there are two types of branches in our model: the main branch to provide the eventual detection results and the auxiliary branches to aid in training.
The main branch employs an embedding-level approach while the auxiliary branches employ an instance-level approach with different pooling methods.
All the branches share the same high-level feature output by the feature encoder and predict the clip-level probability separately.
Since all the branches are independent of each other, they optimize the feature encoder according to their unique learning purpose respectively. In this way, the feature encoder is able to obtain different characteristics due to the efforts of different branches.
The losses of all the branches are calculated by cross entropy loss function with the clip-level ground truth
\begin{equation}
    L = - \sum_c(y_c log(\hat{\mathbf{P}}({y_c}|\mathbf{x})) + (1-y_c) log(1-\hat{\mathbf{P}}({y_c}|\mathbf{x})))
\end{equation}
and utilized to update the feature encoder synchronously.
\begin{equation}
    L_{total} = \alpha L_{main} +\sum{ \beta L_{auxiliary}}
\end{equation}
During training, we set $\alpha$ to be 1.0, and $\beta$ to be 0.5.

\begin{figure}[t]
\begin{minipage}{0.51\linewidth}
  \centering
  \centerline{\includegraphics[width=0.7\linewidth]{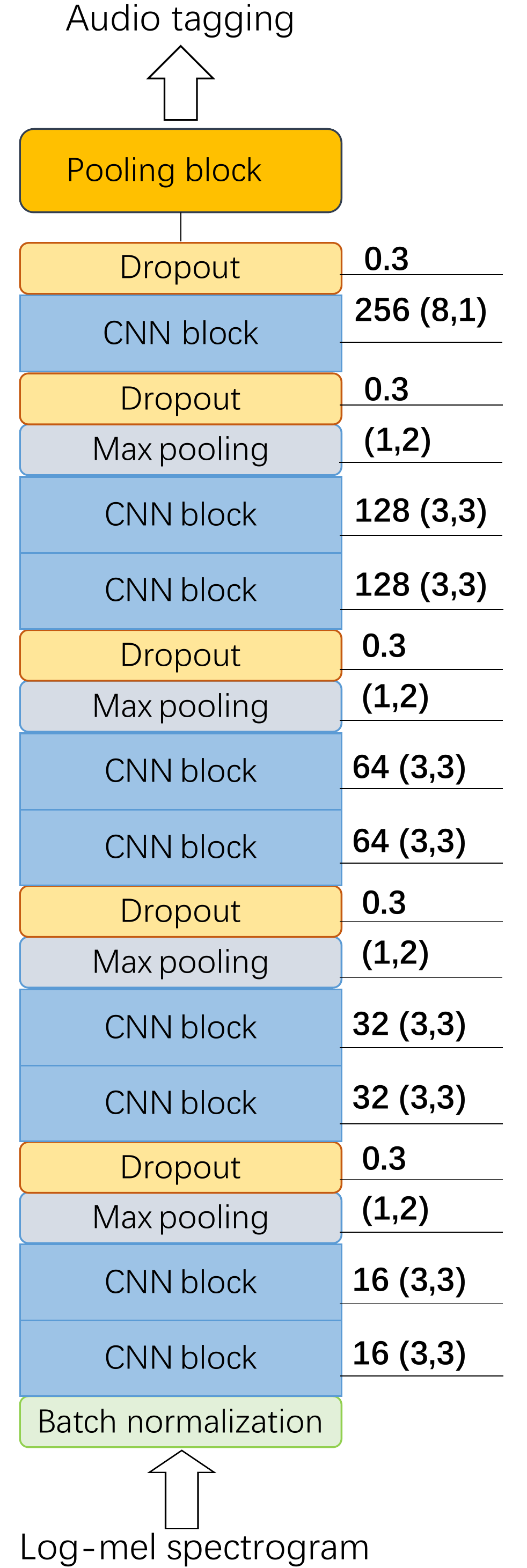}}
  \centerline{(a) URBAN-SED}\medskip
\end{minipage}
\hfill
\begin{minipage}{0.46\linewidth}

\begin{minipage}{0.8\linewidth}
  \centerline{\includegraphics[width=\linewidth]{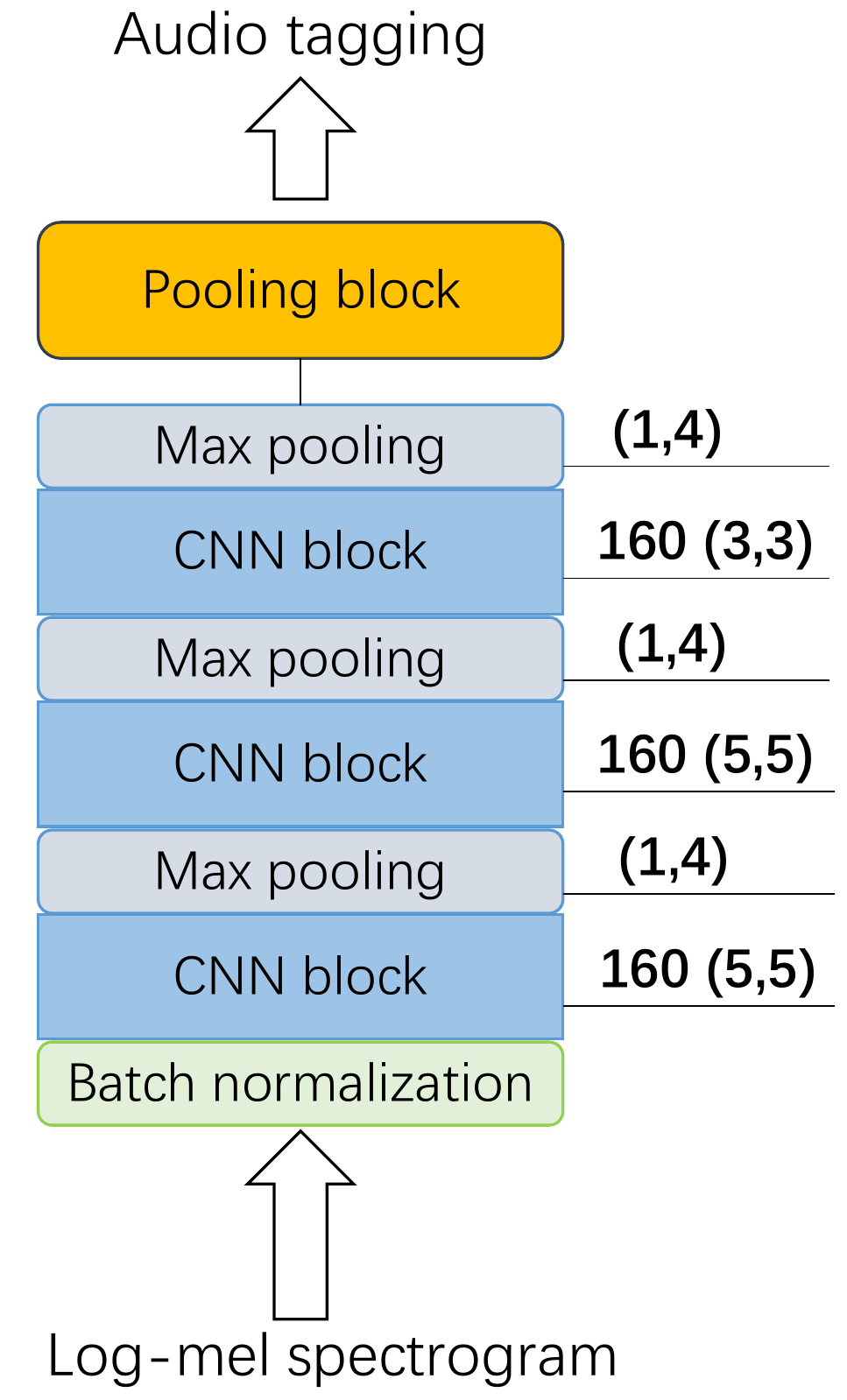}}
  \centerline{(b) DCASE 2018 Task 4}\medskip
\end{minipage}

\begin{minipage}{1.0\linewidth}
  \centering
  \centerline{\includegraphics[width=\linewidth]{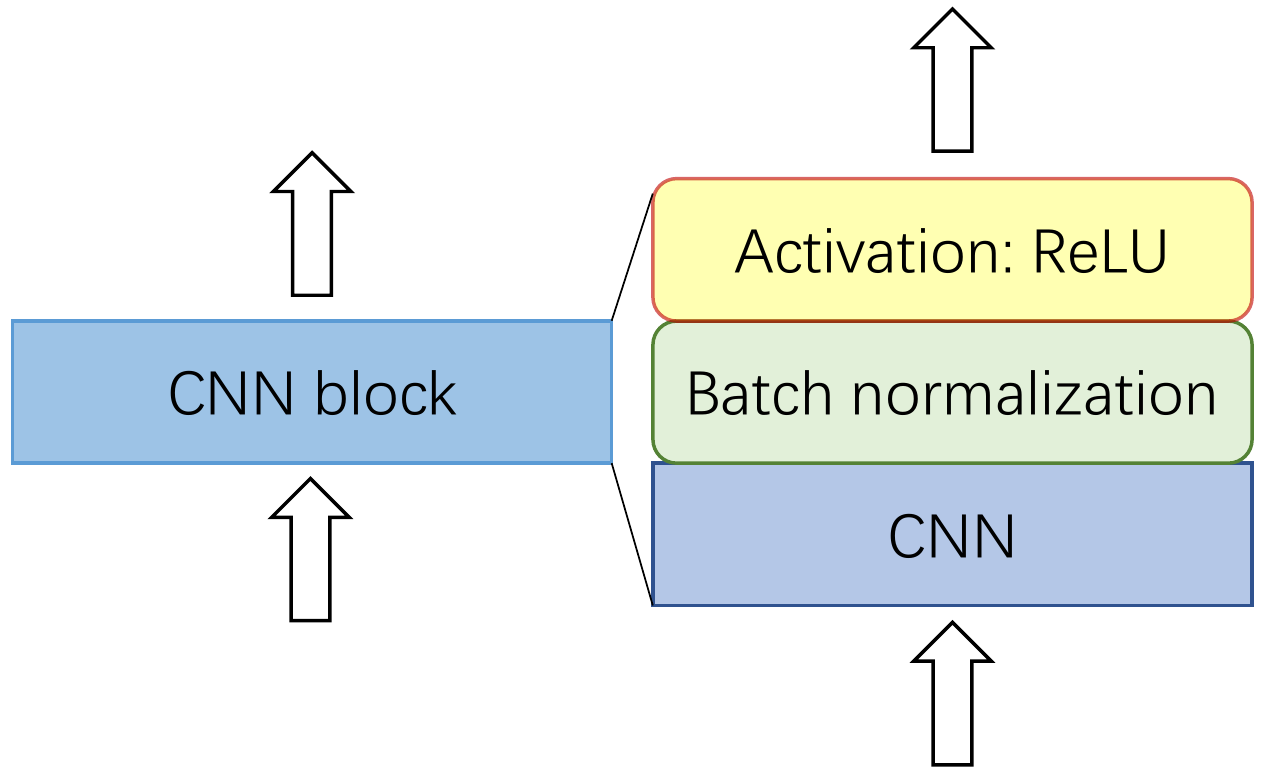}}
  \centerline{(c) CNN Block}\medskip
\end{minipage}
\end{minipage}
 \vskip -0.1in
\caption{The model architectures}
\vskip -0.2in
\label{fig3}
\end{figure}

Furthermore, we employ the CNN framework illustrated in Figure~\ref{fig3} as our feature encoder.
In this figure, the pooling block contains a pooling module and a classifier.

\section{Experiment}
\label{sec:experiment}

\subsection{Dataset}
\label{ssec:dataset}

\subsubsection{DCASE 2018 Task 4}
\label{sssec:DCASE 2018}
The DCASE 2018 Task 4 dataset \cite{serizel2018large}
contains 10 domestic sound event classes.
The dataset is divided into 4 parts: a weakly-labeled training set (1578 clips), an unlabeled training set (54411 clips), a strongly-labeled validation set (288 clips) and a strongly-labeled  test set (880 clips). We take the weakly-labeled set as our training set.

\subsubsection{URBAN-SED}
\label{sssec:URBAN-SED}
The URBAN-SED dataset is generated by the Scaper soundscape synthesis library \cite{salamon2017scaper}. The dataset containing 10 sound event classes is divided into 3 parts: a strongly-labeled training set (6000 clips), a strongly-labeled validation set (2000 clips) and a strongly-labeled test set (2000 clips). We only utilize the weak annotations of the training set.

\subsection{Pre-processing and post-processing}
\label{ssec:pre-post}
We employ 64 log-mel bank magnitudes extracting from 40ms frames with 50\% overlap. In this way, all the 10-second audio clips are converted to feature vectors of 500 frames. For post-processing, all the frame-level predictions are smoothed by a median filter, of which the window sizes are adaptive to event categories.

\subsection{Training and evaluation}
\label{sec:training}

The feature encoder of the DCASE 2018 Task 4 model consists of 3 CNN blocks, each of which contains a convolution layer, a batch normalization layer and a ReLU activation layer. For the URBAN-SED dataset,
the feature encoder consists of 9 CNN blocks with dropout because URBAN-SED dataset is much larger than the DCASE 2018 Task 4 dataset.
For embeding-level ATP, we set scale factor $d$ as the number of feature dimension (160) divided by 2.5 for the DCASE 2018 Task 4 model and as the number of feature dimension (1024) divided by 3.0 for the URBAN-SED model.

We report the event-based marco F1 score \cite{mesaros2016metrics} on DCASE 2018 Task 4 and the segment-based marco F1 score \cite{mesaros2016metrics} on URBAN-SED. All the experiments are repeated 40 times with random initiation and we report both the average result and the best result of each model.

\subsection{Results}

\begin{table}[t]
\vskip -0.08in
  \caption{The event-based F1 score for DCASE 2018 Task 4}
  \vskip 0.1in
  \label{table2}
  \centering
\begin{tabular}{lcc}
\toprule
\textbf{Model}
&\textbf{\textbf{Average F1}}
&
\textbf{Best F1}\\
\midrule
The $\mathbf{1^{st}}$ place&-&$0.324$\\
\hline
E-GMP&$ 0.209\pm 0.0168$ & $ 0.243 $\\
E-GMP + I-GMP&$0.225\pm 0.0151$ & $ 0.256 $\\
E-GMP + I-GAP&$0.242\pm 0.0158$ & $ 0.279 $\\
E-GMP + I-GAP + I-GMP&$0.236 \pm 0.0194 $ & $ 0.267 $\\
\hline
E-GAP&$0.208 \pm 0.0086 $ & $ 0.227 $\\
E-GAP + I-GAP&$0.210 \pm 0.0084 $ & $ 0.225  $\\
E-GAP + I-GMP&$0.213 \pm 0.0107 $ & $ 0.232  $\\
E-GAP + I-GAP + I-GMP&$0.217 \pm 0.0081 $ & $ 0.237 $\\
\hline
E-ATP&$0.310 \pm 0.0164$ & $ 0.335 $\\
E-ATP + I-GMP&$0.315 \pm0.0160 $ & $ 0.343 $\\
E-ATP + I-GAP&$0.311 \pm 0.0143$ & $ 0.343 $\\
E-ATP + I-GAP + I-GMP&$ \mathbf{0.320\pm 0.0141}$ & $\mathbf{0.346} $\\
\bottomrule
\end{tabular}
\vskip -0.15in
\end{table}

\label{sec:result_and_analyse}
Experimental results are shown in Table \ref{table2} and \ref{table3}, where E-* denotes the embedding-level
approach and I-* denotes the instance-level approach. E-ATP + I-GAP + I-GMP achieves the best average performance with event-based F1 scores of 0.320 on the DCASE 2018 Task 4 dataset and segment-based F1 scores of 0.590 on the URBAN-SED dataset. The best result of E-ATP + I-GAP + I-GMP on the DCASE 2018 Task 4 dataset achieves 0.346 which outperforms the first place \cite{jiakai2018mean} in the challenge by 2.2 percentage points, while that on the URBAN-SED dataset achieves 0.616, which is merely 3.1 percentage points behind the SOTA model \cite{martin2019sound} trained with strong annotations.

\textbf{More different learning purposes lead to better results}
We find that if the auxiliary branch utilizes the different pooling methods from the main branch, good performance can be achieved.
As shown in Table~\ref{table2} and~\ref{table3}, the model with branches utilizing different pooling methods tends to perform better than that with branches utilizing the same pooling methods and that with only one main branch. More specifically, E-GMP + I-GAP performs better than E-GMP + I-GMP and E-GMP. We argue that this is because that both E-GMP and I-GMP use a small subset (due to global max pooling) of the feature while the E-GAP and I-GAP use the whole feature, so the characteristic of feature that  E-GMP focuses on is more close to that of I-GMP than I-GAP. Similar to multi-task learning where the difference among multiple tasks reduce the risk of making feature encoder overfit one task,
the differences between multiple learning purposes of multiple branches make the feature encoder be more general
and hard to overfit clip-level or frame-level characteristic and improve the combination performance of audio tagging and event boundary detection.
The result that E-GAP + I-GMP performs better than the E-GAP + I-GAP and E-GAP further proves this explanation. Besides, the result of E-GAP + I-GAP is worse than E-GAP in URBAN-SED experiments. We argue that if the auxiliary branch employs the same kind of pooling method as the main branch, the performance of the model might not be improved because the learning purposes of the two branches are not different enough to prevent the feature encoder from overfitting.

\begin{table}[t]
\vskip -0.08in
  \caption{The segment-based F1 score for URBAN-SED}
  \vskip 0.1in
  \label{table3}
  \centering
\begin{tabular}{lcc}
\toprule
\textbf{Model}
&\textbf{\textbf{Average F1}}
&
\textbf{Best F1}\\
\midrule
\multicolumn{2}{l}{Adaptive pooling (weakly-labeled)\cite{mcfee2018adaptive}}&$0.533$\\
Supervised SED\cite{martin2019sound}&-& $0.647  $\\
\hline
E-GMP&$0.528 \pm 0.0198 $ & $ 0.557 $\\
E-GMP + I-GMP&$0.552\pm0.0113 $ & $ 0.578 $\\
E-GMP + I-GAP&$ 0.557\pm 0.0138 $ & $ 0.593 $\\
E-GMP + I-GAP + I-GMP&$ 0.570\pm 0.0091$ & $ 0.592 $\\
\hline
E-GAP&$ 0.545\pm0.0112 $ & $  0.569$\\
E-GAP + I-GAP&$0.531 \pm 0.0134$ & $0.563 $\\
E-GAP + I-GMP&$ 0.552\pm0.0087 $ & $ 0.571 $\\
E-GAP + I-GAP + I-GMP&$ 0.547 \pm 0.0111$ & $ 0.574 $\\
\hline
E-ATP&$0.581 \pm 0.0102$ & $ 0.599 $\\
E-ATP + I-GMP&$0.582 \pm0.0116 $ & $ 0.605 $\\
E-ATP + I-GAP&$0.585 \pm 0.0109 $ & $ 0.600 $\\
E-ATP + I-GAP + I-GMP&$\mathbf{ 0.590}\pm \mathbf{0.0104}$  & $ \mathbf{0.616} $\\
\bottomrule
\end{tabular}
\vskip -0.15in
\end{table}

\textbf{More auxiliary branches leads to better results}
As shown in Table~\ref{table2} and~\ref{table3}, E-ATP + I-GAP + I-GMP, which employs two different auxiliary branches, performs better than E-ATP + I-GAP and E-ATP + I-GMP.
Since the ATP chooses some frame-level features to train the model, the difference between the ATP and GMP or ATP and GAP is not as large as the difference between GMP and GAP. As a result, the improvement of E-ATP + I-GMP or E-ATP + I-GAP is limited. However, if the two auxiliary branches are both added, the difference between main branch and auxiliary branches can be much larger so that improves the performance of the model.

Therefore, more auxiliary branches with different pooling methods can further increase the difference between the main branch and the auxiliary branches so that further reduce the risk of overfitting.

\section{Conclusions}
\label{sec:conclusion}
In this paper, we propose a multi-branch learning method for weakly-supervised SED.
It consists of a feature encoder and multiple branches which share the same feature encoder. The multiple branches with different learning purposes are implemented by combinations of different MIL strategies including the instance-level and embedding-level approach, and different pooling methods. As a result, the feature encoder is expected to have both frame-level and clip-level characteristics so that be prevented from ovefitting any one of the two characteristics. The proposed method achieves an event-based F1 score of 34.6\% on the DCASE 2018 Task 4 dataset and outperforms the first place model by 2.2 percentage points.

\section{ACKNOWLEDGMENT}
This work is partly supported by Beijing Natural Science Foundation (4172058).

\vfill\pagebreak

\bibliographystyle{IEEEbib}
\bibliography{ms}

\end{document}